# S66 Noncovalent Interactions Benchmark Re-Examined: Composite Localized Coupled Cluster Approaches

Emmanouil Semidalas,[1] Golokesh Santra,[1] Nisha Mehta,[1] and Jan M.L. Martin,[1, a)]

[1]*Dept. of Molecular Chemistry and Materials Science, Weizmann Institute of Science, 7610001 Reḥovot, Israel*
[a)] *Corresponding author: gershom@weizmann.ac.il*

**Abstract.** The S66 benchmark dataset for noncovalent interactions (NCIs) is studied through localized coupled-cluster methods and general LNO-CCSD(T)-based composite schemes. Very small root-mean-square deviations (≤ 0.05 kcal/mol) for the low-cost composite approaches from the SILVER reference interaction energies of S66 indicate that one can safely avoid carrying out the largest basis set calculations with veryVeryTight thresholds; instead, additivity corrections in smaller basis sets can be applied. Interestingly, the counterpoise corrections do not have an appreciable effect on the composite schemes. These findings may prove useful for intermolecular and intramolecular NCIs of larger systems.

## INTRODUCTION

Noncovalent interactions (NCIs) are reversible interactions that play an important role in self-assembly,[1–3] design of new functional materials,[4,5] and biological phenomena.[6,7] These interactions include, but are not limited to, hydrogen,[8] chalcogen,[9] and pnictogen bondings,[10] ion-π interactions,[11,12] π-π stacking,[13] hydrophobic effects,[13] etc. The inherent difficulty[14] of accurately measuring NCI energies is one of the main reasons accurate wavefunction *ab initio* methods are employed for this purpose.[15]

For reasons of computational efficiency, NCI studies are increasingly being carried out through localized coupled-cluster (CC) methods, such as the PNO-LCCSD(T) (pair natural orbital localized coupled cluster with all singles and doubles and quasiperturbative triples) approach of the Werner group,[16] the DLPNO-CCSD(T) (domain localized pair natural orbital CCSD(T)) approach of the Neese group,[17,18] and the Budapest LNO-CCSD(T) (localized natural orbital CCSD(T)) method.[19–21] A few examples are recent studies of the $(H_2O)_{20}$ cages through PNO-LCCSD(T)-F12b,[22] (the F12b suffix refers to explicit correlation[23]) the L7 set[24] of seven large noncovalent dimers using LNO-CCSD(T),[25] and the large and chemically diverse GMTKN55 (general main-group thermochemistry, kinetics, and noncovalent interactions, 55 problem subsets) database[26] through DLPNO-CCSD(T).[27] An important feature of localized CC methods is their asymptotic linear scaling behavior with system size compared to the $N^7$ scaling of the 'gold-standard' CCSD(T). Furthermore, from recent benchmark studies on the highly delocalized polypyrroles[28] and the mechanisms of hydroarylation catalyzed by Ru(II) complexes[29] it became apparent that LNO-CCSD(T) is considerably more resilient toward nondynamical correlation than PNO-LCCSD(T) or DLPNO-CCSD(T).

One price paid for the gentle cost scaling of localized coupled-cluster methods is the introduction of a plethora of cutoffs that arguably represent an empiricism of precision. Generally, tuned fixed combinations of these are presented to the user, e.g., for DLPNO-CCSD(T) one has LoosePNO, NormalPNO, and TightPNO (see Table 1 in Ref.[30] for detailed definitions), and similarly for LNO-CCSD(T) (see Table 1 in Ref.[21]) one has Normal, Tight, vTight, and vvTight. Therefore, one way to estimate the effects of these thresholds is to directly compare the errors for interaction energies of larger systems to canonical CCSD(T) or explicitly correlated CCSD(T)-F12b or CCSD(T)(F12*).[31–33]

In the present work, we investigate the performance of LNO-CCSD(T) for the noncovalent interactions (NCIs) in Hobza's S66 benchmark,[34] using our own revised reference values.[35] The S66 dataset consists of 66 dimers representative of NCIs between biomolecular building blocks and can be subdivided into four different types of interactions: hydrogen bonds, π-stacking, London dispersion complexes, and mixed-influences. In a recent study, Kállay and Nagy[21] employed LNO-CCSD(T)/haV{T,Q}Z, i.e., extrapolating from haVTZ and haVQZ basis sets, and obtained MAEs (mean absolute errors) of 0.33, 0.18, 0.05, and 0.04 kcal/mol, respectively, for Loose, Normal, Tight, and vTight thresholds. The reference were the values labeled "silver [standard]" in Ref.[35], which were obtained as MP2-F12/aV{T,Q}Z-F12 half-CP + [CCSD(F12*)–MP2-F12]/ aVTZ-F12 half-CP + [CCSD(T)–CCSD]/haV{D,T}Z half-CP. ("Half-CP" refers to the average of counterpoise-corrected and uncorrected results.) S66 is one component of the GMTKN55 benchmark suite;[26] in previous studies[36,37] on composite wavefunction theory methods refitted against GMTKN55, we did consider substituting DLPNO-CCSD(T) for CCSD(T) in the post-MP2 correction, but we focused on best overall performance for GMTKN55, not most accurate performance for NCIs.

## COMPUTATIONAL DETAILS

For all single-point energy calculations, we have used the MRCC program suite,[38] version 2020. All the LNO-CCSD(T) calculations were done using correlation consistent aug-cc-pV$n$Z (n = T, Q, 5) basis sets for all nonhydrogen atoms[39,40] and cc-pVnZ for hydrogen atoms; this is denoted haVnZ in this manuscript. The corresponding RI fitting basis sets haVnZ-RI (n = T, Q, 5) were also employed.[41,42] The Boys-Bernardi counterpoise (CP) corrections[43] were applied, and we also considered the average CP or 'half-CP' in the same way as reported previously.[44–46]

A two-point extrapolation was carried out for the LNO-CCSD(T) correlation energies using the following expression from Halkier $et\ al.$[47], $E_{CBS} = E_L + (E_L - E_{L-1})/[\left(\frac{L}{L-1}\right)^\alpha - 1]$, where L refers to the basis set cardinal number and, like in W1 and W2 theories,[48] α=3 for L={Q,5} and α=3.22 for L={T,Q}. The SCF energies were also extrapolated from two basis sets with the equation $E_\infty - E_L = \frac{(L+1)(E_L - E_{L-1})}{L(exp[(\gamma\sqrt{L} - \sqrt{L-1})]-1)}$. The effective exponents $\gamma$ were obtained from Table 1 in Ref.[49]; they are 6.57 for the haV{T,Q}Z, and 9.03 for the haV{Q,5}Z extrapolation.

Next, the optimal values for the linear combination coefficients of the composite LNO-CCSD(T)-based schemes were found after minimizing the root mean square deviations (RMSDs) with respect to the SILVER reference. Full raw data can be found in the ESI at the DOI http://doi.org/10.34933/wis.000397

## RESULTS AND DISCUSSION

TABLE 1. RMSD (kcal/mol) relative to the "silver standard" reference interaction energies[35] of S66 for LNO-CCSD(T) with various basis sets, accuracy thresholds, and counterpoise (CP) corrections. The notation {m,n} refers to extrapolation from haVmZ and haVnZ basis sets

| Basis Set | Threshold | No CP | CP | Half-CP |
|---|---|---|---|---|
| haVTZ | Normal | 0.727 | 0.422 | 0.455 |
| haVQZ | Normal | 0.386 | 0.275 | 0.309 |
| haV5Z | Normal | 0.291 | 0.232 | 0.257 |
| haV{T,Q}Z | Normal | 0.215 | 0.256 | 0.233 |
| haV{Q,5}Z | Normal | 0.232 | 0.212 | 0.218 |
| haVTZ | Tight | 0.581 | 0.380 | 0.297 |
| haVQZ | Tight | 0.217 | 0.170 | 0.141 |
| haV5Z | Tight | 0.114 | 0.108 | 0.096 |
| haV{T,Q}Z | Tight | 0.060 | 0.076 | 0.062 |
| haV{Q,5}Z | Tight | 0.075 | 0.069 | 0.069 |
| haVTZ | vTight | 0.556 | 0.356 | 0.246 |
| haVQZ | vTight | 0.187 | 0.133 | 0.090 |
| haV5Z | vTight | 0.071 | 0.073 | 0.048 |
| haV{T,Q}Z | vTight | 0.039 | 0.033 | 0.028 |
| haV{Q,5}Z | vTight | 0.041 | 0.035 | 0.033 |
| haVTZ | vvTight | 0.563 | 0.336 | 0.243 |
| RMSD from canonical CCSD(T)/haVTZ data taken from ESI of Ref.[35] | | | | |
| haVTZ | Normal | 0.218 | 0.270 | 0.238 |
| haVTZ | Tight | 0.070 | 0.109 | 0.079 |
| haVTZ | vTight | 0.030 | 0.059 | 0.026 |
| haVTZ | vvTight | 0.018 | 0.062 | 0.031 |

The first research question we investigate is the effects of using basis sets larger than haVQZ in LNO-CCSD(T) for predicting the S66 interaction energetics. Table 1 lists the relevant RMSDs for various basis sets, LNO-CCSD(T) thresholds, and counterpoise (CP) corrections against the SILVER reference.

The counterpoise-uncorrected ("raw") results indicate a consistent improvement in accuracy with increasing basis set size and tighter thresholds; the best overall "raw" result is obtained for LNO-CCSD(T,vTight)/haV5Z (RMSD = 0.071 kcal/mol), which is very close to the corresponding CP corrected case. With Normal thresholds and either haVTZ, haVQZ or haV5Z basis sets, the full-CP correction yields RMSD improvements of 0.305, 0.111, and 0.059 kcal/mol, respectively. However, these deviations from raw narrow when thresholds are tightened (see Table 1). Beyond the Normal threshold, half-CP seems to be more beneficial than full CP. As expected, the lowest RMSD (0.048 kcal/mol) is found with the haV5Z basis set and vTight threshold. Again, with vTight, reducing the size of the basis set leads to an RMSD deterioration by just 0.042 kcal/mol for haVQZ, but 0.198 kcal/mol for haVTZ. We note a similar trend for Tight thresholds; a two-fold improvement in RMSD from haVTZ to haVQZ, and an almost three-fold improvement from haVTZ to haVQZ. It has been established for both orbital-based CCSD(T)[44] and for explicitly correlated methods[35] that half-CP tends to converge more smoothly to the CBS limit than either raw or full CP.

For haVTZ, we have canonical results available.[35] RMSD for raw CCSD(T) decays from 0.22 kcal/mol for Normal to 0.07 kcal/mol for Tight to 0.03 kcal/mol for vTight to less than 0.02 kcal/mol for vvTight. The convergence of RMSD for CP and half-CP is more erratic for tighter settings.

Next, we consider the effects of complete basis set (CBS) extrapolation. As expected, the two-point extrapolations significantly improve statistics, and the errors converge more smoothly towards the CBS limit, as long as tighter thresholds are being used. The lowest RMSD error (0.028 kcal/mol) is found for LNO-CCSD(T,vTight)/haV{T,Q}Z half-CP. Also, with half-CP, an haV{T,Q}Z extrapolation with Tight threshold outperforms all single basis set LNO-CCSD(T) methods, except the most expensive LNO-CCSD(T,vTight)/haV5Z (RMSD = 0.048 kcal/mol). Interestingly enough, with "raw" the lower-cost haV{T,Q}Z extrapolation always marginally outperforms the more expensive haV{Q,5}Z; the same is observed for Tight and vTight with half-CP. This is all likely fortuitous, as we estimate the uncertainty for the SILVER reference values[35] to be about 0.02 kcal/mol, i.e. 3 times the RMS difference with GOLD reported there[35] for a small subset. However, since accuracy is significantly improved, we could further study the effects of additional corrections to avoid expensive vvTight calculations with larger basis sets.

Now, let us focus on LNO-CCSD(T)-based composite schemes, here denoted cLNO. In these schemes, a two-point extrapolation is carried out between a 'small' and 'large' basis set with the same thresholds, and the effect of further tightening is then estimated as a scaled additivity correction in a relatively small basis set (see Table 2). Without a CP correction, the lower-cost Tight{T,Q} + 0.54[vTight – Tight]/T composite scheme offers accuracy comparable to the costlier LNO-CCSD(T,vTight)/haV5Z method. The best result is obtained by a three-tier Tight{Q,5}+ 3.65[vvTight – vTight]/T + 0.11[vTight – Tight]/Q scheme (RMSD = 0.042 kcal/mol), which is in fact comparable to vTight/haV{Q,5}Z method (RMSD = 0.041 kcal/mol).

Next, what are the effects of employing full counterpoise? For Normal{T,Q} + 1.02[vTight – Normal]/T, performance deteriorated significantly (~0.04 kcal/mol) when compared to "raw". The three-tier composite scheme is another example where using full CP increases RMSD by only 0.01 kcal/mol. Apart from these two, using CP correction improves RMSD statistics for all other cLNO methods. Interestingly enough, low-cost Tight{T,Q} + 0.70[vTight – Tight]/T with full-CP correction appears to be the best non-haV5Z option with RMSD = 0.050 kcal/mol.

**TABLE 2.** RMSD (kcal/mol) and linear combination coefficients for the LNO-CCSD(T)-based composite wavefunction methods (cLNO) against the SILVER standard reference interaction energies of S66 using various basis sets, accuracy thresholds, and CP corrections.

| cLNO methods | Raw | | | CP | | | Half-CP | | |
|---|---|---|---|---|---|---|---|---|---|
| | RMSD (kcal/mol) | coefficients | | RMSD (kcal/mol) | coefficients | | RMSD (kcal/mol) | coefficients | |
| | | $c_1$ | $c_2$ | | $c_1$ | $c_2$ | | $c_1$ | $c_2$ |
| Normal{T,Q} + $c_1$[vTight – Normal]/T | 0.055 | 0.86 | – | 0.094 | 1.02 | – | 0.063 | 0.95 | – |
| Normal{Q,5} + $c_1$[vTight – Normal]/T | 0.094 | 0.96 | – | 0.081 | 0.84 | – | 0.072 | 0.90 | – |
| Tight{T,Q} + $c_1$[vTight – Tight]/T | 0.063 | 0.54 | – | 0.050 | 0.70 | – | 0.042 | 0.61 | – |
| Tight{T,Q} + $c_1$[vvTight – Tight]/T | 0.057 | 0.60 | – | 0.049 | 0.63 | – | 0.035 | 0.60 | – |
| Tight{Q,5}+ $c_1$[vTight – Tight]/T | 0.075 | 0.98 | – | 0.061 | 0.40 | – | 0.061 | 0.63 | – |
| Tight{Q,5}+ $c_1$[vTight – Tight]/Q | 0.077 | 0.88 | – | 0.060 | 0.45 | – | 0.059 | 0.70 | – |
| Tight{Q,5}+ $c_1$[vvTight – Tight]/T | 0.062 | 1.00 | – | 0.057 | 0.44 | – | 0.052 | 0.70 | – |
| Tight{Q,5}+ $c_1$[vvTight – vTight]/T + $c_2$[vTight – Tight]/Q | 0.042 | 3.65 | 0.11 | 0.054 | 1.31 | 0.26 | 0.037 | 2.38 | 0.27 |

Except for the Normal{T,Q} + 0.95[vTight – Normal]/T, half-CP corrected energies offer a performance improvement for all cLNO schemes over either raw or full-CP across the board. For the lowest-cost approach, Normal{T,Q} + 0.95[vTight – Normal]/T, we obtain RMSD = 0.063 kcal/mol, which is marginally better than costlier LNO-CCSD(T,Tight)/haV{Q,5}Z (see Table 1). With RMSD = 0.035 kcal/mol, Tight{T,Q} + 0.60[vvTight – Tight]/T, can offer accuracy comparable with LNO-CCSD(T,vTight)/haV{T,Q}Z and LNO-CCSD(T,vTight)/haV{Q,5}Z. Hence, we can argue that these composite schemes can safely reproduce the SILVER reference energies at a tiny fraction of their (formidable) computational cost.

We note that in Section 6.1 of Ref.[21] an estimate for "tighter" contributions is presented that corresponds to $E_{vTight} \approx [E_{Tight}+(E_{Tight}-E_{Normal})/2] \pm (E_{Tight}-E_{Normal})/2$. Recursively applying this expression to go to ever more "v"s eventually converges to $2E_{Tight}-E_{Normal}=E_{Normal}+2(E_{Tight}-E_{Normal})$, which may rationalize, after a fashion, the $c_1>1$ in the last entry of Table 2.

Figure 1 presents, for different basis sets and accuracy thresholds, the total wall times for the parallel-displaced benzene dimer. As can be seen, all else being equal, each notch by which we tighten thresholds roughly triples computational cost. The most expensive calculation, LNO-CCSD(T, vvTight)/haVQZ, took nearly 3 days wall time.

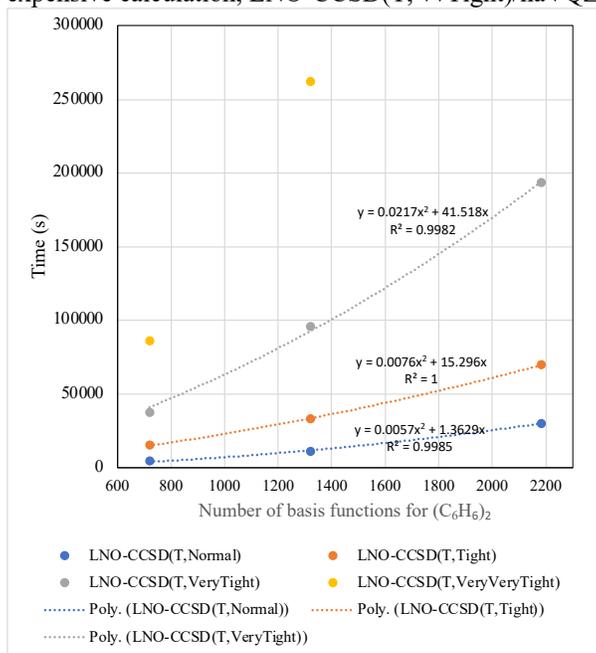

**FIGURE 1.** Total wall time (s) for the parallel displaced benzene-benzene dimer (system 24 in S66) with LNO-CCSD(T) using different basis sets and accuracy thresholds on two 8-core Intel Xeon E5-2630 v3 CPUs (2.40 GHz).

While the proposed cLNO composite schemes offer high accuracy, the question is whether they realize any gains over direct calculation at the highest tolerances? One indication is the total computational cost of selected cLNOs over plain vTight LNO-CCSD(T) with the largest basis set: again, we turn to the parallel displaced benzene-benzene dimer for an estimate. With half-CP, the lowest-cost cLNO, Normal{T,Q} + 0.95[vTight – Normal]/T (RMSD = 0.063 kcal/mol), is 1.8 times less expensive than [vTight]/QZ and 3.6 times less expensive than [vTight]/5Z.

Also, one of the most accurate composite schemes, Tight{T,Q} + 0.61[vTight – Tight]/T half CP (RMSD = 0.042 kcal/mol) is 1.5 times more expensive than [vTight]/QZ. Applying vvTight thresholds for the smallest basis set is only 68% costlier than the [vvTight]/TZ step in isolation, but 1.8 and 3.7 times more affordable than [vvTight]/QZ and [vvTight]/5Z, respectively. Hence, the proposed cLNOs could be employed to reach similar or higher accuracy at a cost that is several times less than [vTight or vvTight]/haV$n$Z with $n$ = Q or 5. This finding is useful for larger systems, where vTight or vvTight LNO-CCSD(T) calculations with larger $n$-tuple ζ basis sets might still be intractable due to CPU time and resource constraints.

# CONCLUSIONS

To sum up, the focus was on LNO-CCSD(T)-based composite methodologies, and we investigated the interaction energies of the S66 database with respect to the "silver standard" reference from Ref.[35].

Tight thresholds in LNO-CCSD(T) with one haV{T,Q}Z or haV{Q,5}Z extrapolation, plus an additive correction based on tighter thresholds but smaller basis sets, are effective and can help us avoid the much more expensive large-basis set calculations with vvTight thresholds. Also, these schemes make CP corrections nearly redundant since statistical improvements are comparable to the residual uncertainty in the reference values. The latter is useful to know in intramolecular interactions where CP is awkward to apply.

The cost-benefit ratio of these cLNO schemes is appreciable, with very similar accuracy to LNO-CCSD(T) with vTight or vvTight thresholds, and thus we recommend them for intra- and intermolecular NCIs of larger systems.

# ACKNOWLEDGMENTS

This research was supported by the Israel Science Foundation (grant 1969/20) and by the Minerva Foundation (grant 2020/05). G.S. acknowledges a fellowship from the Feinberg Graduate School (Weizmann Institute). The work of E.S. on this scientific paper was supported by the Onassis Foundation—Scholarship ID: FZP 052-2/2021-2022.